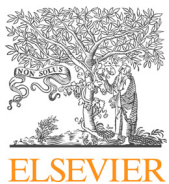
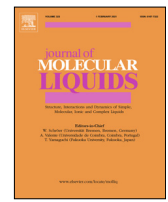

# Molecular simulation of methane hydrate growth confined into a silica pore

Ángel M. Fernández-Fernández [a], María M. Conde [b], Germán Pérez-Sánchez [c], Martín Pérez-Rodríguez [a], Manuel M. Piñeiro [a],*

[a] CINBIO, Dpto. de Física Aplicada, Univ. de Vigo, 36310, Spain
[b] Dpto. de Ingeniería Química Industrial y Medio Ambiente, Escuela Técnica Superior de Ingenieros Industriales, Universidad Politécnica de Madrid, 28006, Spain
[c] Departamento de Química, CICECO, Universidade de Aveiro, Portugal

A B S T R A C T

The growth of a methane hydrate seed within a silica slit pore of fixed width has been studied using All-Atom Molecular Dynamics (AA-MD). An AA force field has been used to describe the molecules of the solid silica substrate, with $\alpha$-quartz crystalline structure. The crystallisation of hydrates in confined geometries is not well understood yet, and the objective of this work is to study the hydrate growth inside a silica pore using molecular simulation. Both NVT and NpT ensembles were used in the AA-MD simulations to analyse the hydrate growth from an initial seed. Results showed that the boundary conditions imposed by the nanometric slit pore yielded a hydrate with structural defects, filling the accessible space between the silica walls. The water molecules which were not incorporated to the initial seed hydrate formed a high density water layer trapped between the silica walls and the crystallised hydrate. These results provide an interesting insight into the hydrate crystallisation process in confined geometries, resembling those found in natural hydrate deposits.

## 1. Introduction

Gas hydrates are non-stoichiometric inclusion crystalline solids composed by host water molecules [1]. This structure is stabilised by hydrogen bonds, and leaves cells that may enclathrate small guest molecules. The most ubiquitous gas hydrate is methane hydrate, whose deposits can be widely found in ocean seabeds and also in frozen permafrost soils. Geological explorations and seismic analysis of those emplacements support the affirmation that the abundance of methane trapped into these natural deposits is enough to be considered as a strategic energy source [2–8]. On the other hand, hydrates might also represent an environmental threaten to be seriously considered, since high concentrations of methane in the atmosphere can dramatically affect the global temperature, contributing to the Global Warming even more than $CO_2$ [9]. Thus, eventual leaks from these sources might have a harmful impact in the ongoing climate change process [10]. The total amount of methane trapped as hydrates has been a matter of controversy during the last decades, but even the most conservative lower bounds for these estimations are breathtaking [2,11,12]. Only in the Gulf of Mexico, The Minerals Management Service of the United States estimated that about 190 trillion m$^3$ of methane are hidden as hydrates [13].

This fact has boosted global research efforts, investment and technical development on potential extraction methods for these hydrate deposits. One of the most remarkable progress in the recovery of methane from gas hydrates has been achieved by the government of Japan in the Nankai Trough in 1999. This area was prospected during many years, and it was discovered that the main deposits of methane hydrate appear as turbidic sandy layers, filling the intergranular porosity [14]. The permeability of these reservoirs, confined by layers of impermeable sediments, could be the key to exploit those methane hydrate deposits as a feasible energy resource [15].

Bearing in mind the explained reasons, a deep knowledge of the structure, thermodynamics, and phase equilibria of hydrates under confinement arise as relevant objectives to improve the feasibility of the potential applications proposed. Taking into account the nature of the sediments and rocks where natural hydrates can be found, the influence of a solid substrate in the hydrate nucleation, growth and/or dissociation processes, remains unclear and stand as challenging objectives nowadays. The characterisation of hydrates phase equilibria in confined spaces from a molecular scale can shed light into the complex scenario where hydrates are usually found, under extreme temperature and pressure conditions,

⇑ Corresponding author.
E-mail address: mmpineiro@uvigo.es (M.M. Piñeiro).

mostly unreachable in laboratories. Over the last decades, the progress in computational performance has extended the accessible ranges of time and length scales in the application of molecular simulation techniques. In this context, the theoretical analysis of hydrates using computational approaches has seen remarkable progress, as summarised in the reviews of Varini et al. [16], Barnes and Sum [17], or English and MacElroy [18]. However, these reviews evidence that the number of studies concerning hydrates in the presence of solid substrates or under confinement conditions is still very scarce. It must be noticed that some authors have performed MD calculations including a sand porous media in their simulation boxes [19–21]. In those simulations, two silica walls of $\alpha$-quartz were designed to mimic the natural conditions of hydrate formation in a sandy confined nanopore, and the positions of all atoms of the crystal framework were kept frozen. In that sedimentary environment, the dissociation process of methane hydrate was studied. Thus, the melting of methane hydrate is normally obtained by setting a MD thermostat which allows maintaining a temperature above the triphasic equilibrium temperature.

In previous works, we have used MD simulations to analyse the phase equilibria and stability of methane [22,23] and carbon dioxide [24] hydrates. The interaction among neighbouring guest molecules [25], or the phase equilibrium shift on methane hydrates induced by salinity in oceanic conditions [26] have been also studied using the same approach. *Ab initio* quantum calculations were used as well as an alternative to study guest molecule transitions for type *I* hydrates [27,28].

Considering the interest of the study of confined hydrates, the objective of this work is to provide a nanoscopic scenario regarding the growth of the methane hydrate in a confined geometry. A planar slit pore composed by two silica walls described by an atomistic force field has been considered as a first approach. The description of the solid substrate used was inspired in the previous works carried out by Pérez–Sánchez et al. [29,30], where MD simulations of silica and cetyltrimethylammonium bromide (CTAB) were performed using both atomistic and coarse grain (CG-MD) approaches to reproduce the synthesis of MCM-41 mesoporous materials.

Combining these elements, we have employed AA-MD simulations to describe the growth of a methane hydrate inside a planar silica pore where the solid substrate is described by a fully flexible geometry. This approach allows to analyse how the crystalline hydrate cells are rearranged in this confined space, as well as the behaviour of the interface between the hydrate and the solid substrate. The above scenario was highlighted in the work of Bagherzadeh et al. [21] The authors noted in their results that the arrangement displayed by the water molecules closer to the silica interface, could be an artefact due to the frozen silica wall they used. This is the reason why we decided to model $\alpha$-quartz walls where the positions of all atoms are allowed to move. A complete characterisation of their intra and intermolecular force fields provides flexibility to the confining solid substrate. Under this configuration, we intended to mimic the geological environment where hydrate-bearing sediments are confined.

The manuscript is organized as follows. The next section describes molecular simulation calculation details, including the different force fields used, computational details and description of structural characterization methods employed. The results obtained are discussed in the following section, and finally the main conclusions of the study are summarized.

## 2. Computational section

### 2.1. Molecular modelling

An all-atom force field was selected to describe the slit pore, composed by two parallel silica walls. The total potential energy function consists in the addition of bonded interactions, namely the bond vibrations, bond angles stretching and dihedral torsion, whereas the non bonded interactions include the dispersive Lennard-Jones and the Coulomb terms. Methane was modelled using the all-atom Optimised Potential for Liquid Simulations (OPLS-AA) [31] force field, and the well-known rigid non polarizable TIP4P/Ice [32] model was used for water molecules.

The election of the water molecular model is a key point in the simulation, specially in clathrate hydrate studies. A comprehensive set of water models were employed in the work of Conde and Vega [22] for computing the three phase coexistence line of methane hydrate. Their results showed that TIP4/Ice [32] water model offered the best agreement with experimental data. Miguez et al. [24] performed a similar study for carbon dioxide hydrates and TIP4P/Ice [32] also provided the best results. For that reason, we have employed here the well-known rigid and non-polarizable TIP4P/Ice [32] model, as we did also in previous works [26,33].

The parameters of silica molecules were taken from the work of Azenha et al. [34], while the dispersive Lennard-Jones parameters were obtained from Smith et al. [35]. The crystalline unit cell of $\alpha$-quartz structure was extracted from the American Mineralogist Crystal Structure Database [36]. This unit cell was replicated $8 \times 8 \times 4$ times, thus, the silica block has a dimension of $4.022 \times 3.483 \times 2.265$ nm. Periodic boundary conditions were applied in all directions. The silica surface was built by removing the top silica atoms and hydrogens were attached to the terminal oxygens, ensuring that the net neutral charge is maintained. This procedure was followed by other authors in literature to saturate the surface of amorphous silica [37,38]. Thus, the silica structure contains 2496 atoms, where 1536 are oxygen, 704 silicon and 256 hydrogens.

### 2.2. Simulation details

The initial configuration of the system is shown in Fig. 1. Once the silica block has been built, the Z size of the simulation box is extended to 9.26 nm, leaving an empty space where the fluid water, methane and seed hydrate will be placed. The silica block is then displaced, centering the empty space in the simulation box in the Z direction. Thus, the two silica surfaces perpendicular to this axis constitute now the parallel slit pore where the growing hydrate is confined. Then, in the middle of this nanopore a seed of type *I* methane hydrate is placed. The crystalline seed consists in a hydrate whose unit cell was replicated twice in the three directions of space, yielding a total of eight unit cells. The seed was built with a methane guest occupancy of 100%. This means that there are 368 water molecules in the hydrate seed lattice structure, and 64 methane molecules initially enclathrated. Then, the seed size is $2.4 \times 2.4 \times 2.4$ nm, and it has been initially placed in the pore geometrical centre. The system is completed surrounded by 2272 water molecules randomly placed with an average density corresponding to the thermodynamic conditions considered. Then, 256 additional methane molecules were finally placed in a thin slab between liquid water and the solid silica substrate. Those two layers represent the supply of free methane necessary for the hydrate growth. The total dimension of the simulation box was $4.02 \times 4.02 \times 9.26$ nm. and the distance between both silica walls, representing the pore width, is 7 nm.

The AA-MD simulations were carried out integrating the Newton's equations of motion of all particles employing the leap-frog algorithm [39] and a time step of 2 fs. LINCS [40] algorithms were set up for constraints in all bonds. The Lennard-Jones and the real part of the Coulombic potential were truncated at 1.5 nm. The long-range electrostatic interactions were handled using Ewald sums method [41]. Simulations were performed in the NVT and NpT ensembles, where Nosé-Hoover thermostat [42,43] and Parrinello-Raman barostat were used [44,45] with coupling time constants of 0.2 and 0.5 ps, respectively. Once the simulation box was built, the energy of the system was minimised using a steepest descent algorithm. Then, a first equilibration step of 100 ps was performed in the NVT ensemble while the hydrate seed was kept frozen allowing the water molecules to rearrange around the crystal. Then, the seed was unfrozen and the system was equilibrated for 100 additional ps. Once the system was fully equilibrated, three simulations were performed, two in the NVT ensemble at 260 and 273 K and one in the NpT ensemble at 100 bar and 260 K. These thermodynamic conditions are in the phase space region of the force fields employed where the methane hydrate is stable in bulk conditions [22] since according to the MD determined triphasic equilibrium line, methane hydrate melts at ≈ 285 K when pressure is 100 bar. All simulations were run with GROMACS/2021.1 [46,47] software package.

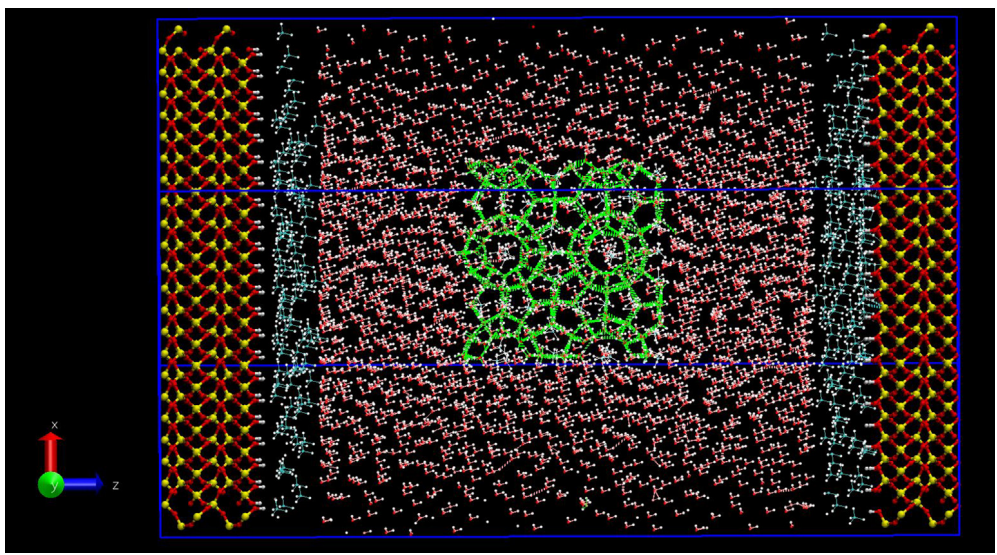

**Fig. 1.** Snapshot of the initial simulation setup. The silica walls are composed of silicon, oxygen and hydrogen atoms, represented in yellow, red and white, respectively. The hydrate seed is placed in the middle of the pore, and the building water molecules are depicted in green. The other water molecules, surrounding the hydrate seed, are shown in red (oxygen atoms) and white (hydrogen atoms). Methane molecules filling the gap between silica wall and liquid water are represented with blue carbon atoms and white hydrogen atoms.

*2.3. Characterisation tools*

The evolution of the initial methane hydrate seed was then monitored along 400 ns, and in each case it was ensured that the system attained a stationary state after this initial period.

Additionally, 100 ns were run as production period to evaluate the structure inside the pore, including the water–silica interface. With the aim to characterise the structure of the methane hydrate inside the nanopore, $F_3$ [48] and $F_4$ [49] order parameters, and hydrogen bond angles, were computed inside the pore in slabs along the Z axis. This procedure allowed to obtain the profile of each magnitude along the pore. The $F_3$ order parameter assesses the tetrahedral environment of each water molecule, which is bonded to other four molecules. It is a three-body structural parameter, and computes the degree of deviation that each oxygen atom has from an perfect tetrahedral network environment.

$$F_3 = \left(\cos\theta_{icj} \mid \cos\theta_{icj} \mid + \cos^2\theta_t\right)^2 \qquad (1)$$

$\theta_{icj}$ is the angle between triplets of oxygen atoms. The oxygen labelled $c$ is the atom of the central water molecule which is compared with other two neighbor oxygen atoms ($i$ and $j$) which belong to those molecules placed within a spherical shell of 0.35 nm. This distance value matches the first minimum in the $O - O$ radial distribution function (RDF) of liquid phase water molecules, and $\theta_t$ = 109.47° is the tetrahedral angle. According to Eq. 1, $F_3$ parameter should vanish when water molecules form a perfect tetrahedral network and increase with the disorder in water structure. Thus, in our simulations, we can distinguish between liquid and solid-like water, but not between ice or hydrate solid phases. Therefore, it is necessary to use an alternative order parameter, in this case $F_4$, which can discern not only between solid and fluid water, but also between hydrate and ice solid phases, as explained by Walsh et al. [50] and Moon et al. [51] The $F_4$ numerical value is a function of the torsion angles between the outermost hydrogen atoms of two water molecules which have their oxygen atoms at a distance below 0.3 nm.

$$F_4 = \cos 3\phi \qquad (2)$$

Where $\phi$ is the torsion angle. According to Eq. 2, the averaged value of this parameter for a liquid water is −0.04, while for hydrate and ice are 0.7 and −0.4, respectively. The only limitation here is that $F_4$ can not distinguish between different hydrate structures.

In our simulations, the hydrogen bond angle between two neighbouring water molecules was also computed. When a hydrogen bond is established, a triangle can be drawn between the two oxygen atoms and the bridging hydrogen. Within this triangle, the $\widehat{O_d H O_a}$ angle has been determined, where $O_d$ is the donor oxygen, covalently bonded to the hydrogen atom, and $O_a$ is the acceptor oxygen in the neighbour water molecule. By solving the triangle, the angle formed by the intramolecular OH bond with the bridging distance between these two intermolecular oxygen atoms has been also determined, and referred as $\widehat{H O_d O_a}$ angle. The geometrical criteria suggested by Mancera et al. [52] has been considered, computing the $\widehat{O_d H O_a}$ angle, while Luzard et al. [53] determine the $\widehat{H O_d O_a}$ angle instead. Both criteria established that the distance between the donor and the acceptor oxygen atoms must be shorter than 0.35 nm. Accordingly, $\widehat{O_d H O_a}$ angle is in the range 130° − 180°, and $\widehat{H O_d O_a}$ must be lower than 30°.

Then, with the aim to study the thin layer of water molecules that will be shown to appear between the silica walls and the hydrate, the residence correlation function $C_R$ [54] can be used,

$$C_R = \frac{O_w(t)\, O_w(0)}{O_w(o)\, O_w(0)} \qquad (3)$$

In this equation $O_w(t)$ is a binary index with values 1 or 0 depending on whether the water molecule including this oxygen atom remains in this layer after a certain $t$ period of time (value 1) or not (value 0). The value of this function is averaged along the simulation trajectories for all molecules in the layer, thus the time dependency shows the proportion of water molecules that have left this layer. It must be noticed that $t$ tends to 0 as water molecules in this layer are exchanged.

## 3. Results and Discussion

Fig. 2 illustrates the result for the NVT simulation at 273 K. The NVT and the NpT production runs at 260 K systems exhibited a quite similar behaviour, thus we focused in the first case to evaluate the methane hydrate growth in the confined space. Additionally, the plots obtained for the other simulations performed have been included in the Appendix, showing equivalent behaviour in all cases. For the sake of clarity, only the representation shown in Fig. 2 is commented in detail. As it can be noticed in Fig. 2, the original hydrate seed has grown up through a crystallisation process, eventually occupying the accessible volume in the pore. It must be highlighted again that the complete system is fully flexible, including the solid silica walls. During the crystallisation process, methane molecules moved away from the hydrophilic silica surface where they were originally located, and were finally enclathrated as guests inside the new growing hydrate cells. Type I hydrate crystalline structure is composed of two different cells, with geometrical shapes of a regular dodecahedron (D type, or $5^{12}$ cell) and a truncated hexagonal trapezohedron (T type, or $5^{12}6^2$). The regular shape of these hydrate cells can be easily identified by a visual inspection in the simulation snapshot shown in Fig. 2, indicating that all the accessible volume has been occupied by this type I hydrate. Nevertheless, it is also clear that the original seed has not maintained its original position, and a rotation along the hydrate main axis was noticed.

The confined hydrate structure exhibits an imperfect structure when compared with unconfined configurations [22,24], exhibiting defects and structure distortion. Most of the original liquid phase water molecules were eventually incorporated into the crystal hydrate sI structure, growing the original methane hydrate seed until the geometrical limit imposed by the silica wall boundaries are reached. The new grown hydrate cells are rotated and somehow distorted to accommodate and fill in the accessible confined space. A visual inspection denotes clearly some of these imperfections, producing that the characteristic linear channels along three dimensions in sI hydrate structure, clearly identified in the initial seed, can be now recognized but appear to be clearly distorted.

Fig. 3 shows the internal energy profiles of the three simulation runs performed. In all of them, a monotonically energy decrease trend was noticed along the simulation, reaching a plateau after 300 ns of simulation time, denoting that the system attained the thermodynamic equilibrium. The obtained energy profile denotes a crystallisation process, as pointed out in many previous works [26,33,22,24,55]. The crystal growth progress was very slow due to the low solubility of methane in water, thus the mass transfer of methane to the water–hydrate interface is the key and the limiting step [1].

The hydrate crystal growth can also be monitored through the evolution of the molecule density profiles across the pore. With this aim, the average density profiles are displayed and compared at the beginning and the end of the NVT (273 K) simulation run in Figs. 4a and b, respectively. At the beginning of the simulation, the hydrate seed was placed in the centre of the nanopore, surrounded by liquid water molecules as it can be noticed in the density profile. Fig. 4a displays also the density profile of methane, denoting how these molecules were initially placed next to the silica walls, and also enclathrated in the hydrate seed in the center of the nanopore. In comparison, the final equilibrium state displayed in Fig. 4b shows two clearly discernible regions inside the pore. The centre of the pore is occupied by an ordered crystalline hydrate slab, where sharp water density peaks appear, This crystallized hudreta region has been labelled as H layer in Fig. 4b. Then, between this central crystalline hydrate region and each silica wall two high density water layers appear, denoted as F layers in Fig. 4b. These adsorbed water F layers are approximately 1.0 – 1.5 nm width.

The adsorbed water layers were formed with the liquid phase water molecules that were unable to join the crystalline structure, yielding two confined water layers between the hydrate and the silica substrate. In addition, a detailed analysis of the nature and structure of these confined adsorbed F layers, each of them roughly 1.5 nm wide, demonstrated that their state was fluid, as shown below. At the end of the simulation, the density profile shown in Fig. 4b denotes how methane molecules have migrated from the region where they were set up in the initial configuration, due to the hydrophilic behavior of the hydroxylated interface of the silica wall. Methane molecules moved through the liquid water region to reach the hydrate-liquid interface. This behaviour contributes to the growth of the type I hydrate cells around them as guest molecules. In fact, all methane molecules present were incorporated to the hydrate structure, and none of them remained close to the silica surface.

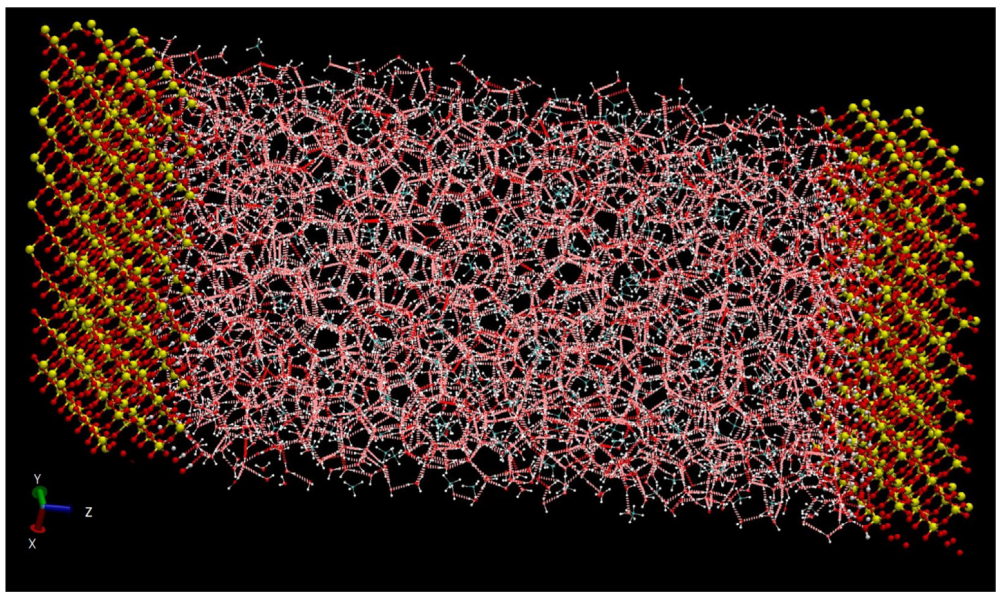

**Fig. 2.** Snapshot of the final state of the NVT simulation at 273 K. All molecules are displayed with the same color code as Fig. 1, and additionally hydrogen bonds are represented with red-whi.te dotted lines.

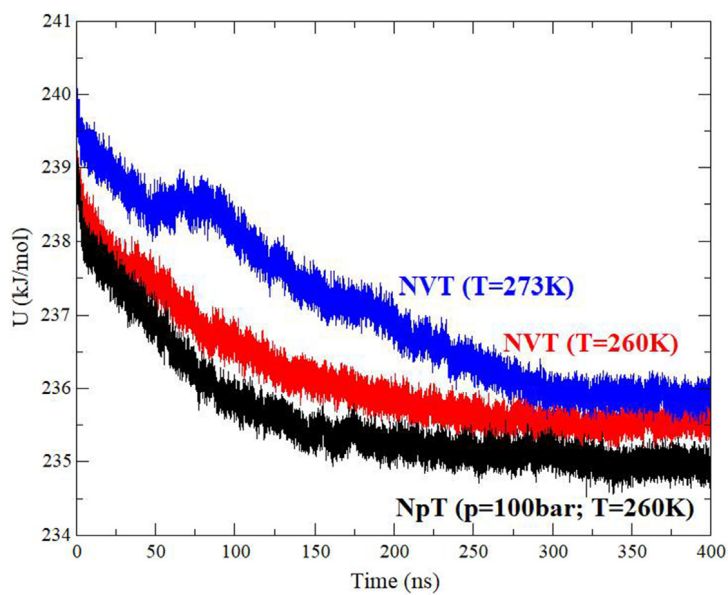

**Fig. 3.** System internal energy profile for the three simulations performed, showing the energy decrease corresponding to the crystallisation process inside the pore. The energy trend in NpT ensemble (p = 100 bar; T = 260 K) is plotted in black. The energy profiles of the simulations performed in the ensemble NVT at 260 K and 273 K are shown in red and blue lines, respectively.

The density profiles shown in Fig. 4 highlight the main differences between the structures in a non confined system with those enclosed between two silica walls, where the water molecules are adsorbed to the hydroxylated silica wall due to its hydrophilic nature. Furthermore, the cells of the structure were rearranged inside this nanopore, where molecules were compelled to accommodate in the confined space and as a consequence, the crystal hydrate showed some imperfections. Then, with the objective to provide a deeper insight into the final system, the $F_3$ and $F_4$ order parameters, and the hydrogen bond angle distributions, were computed to determine the orientation of water molecules in the hydrate structure and also in the vicinity of the silica wall surface. First, the values of such parameters were computed by comparing two systems, a benchmark bulk hydrate system containing an unconfined methane hydrate, and a bulk water system with only liquid water, as described in Table 1. The benchmark bulk hydrate system was set up as follows: The bulk phase of methane hydrate was set twice larger than the seed hydrate previously used. This means that this system contains 736 water and 128 methane molecules, with dimensions 2.4 × 2.4 × 4.8 nm. Three simulations were performed at 260 and 273 K in the NVT, and at 100 bar and 260 K in the NpT ensembles. Then, the benchmark water system was run in a NpT ensemble at 300 and 350 K (1 bar) containing 900 water molecules in a 3.0 × 3.0 × 3.0 nm simulation box as described in Table 1. Simulation setups were the same as previous runs.

The values of computed order parameters $F_3$ and $F_4$, and hydrogen bond angles $\widehat{O_dHO_a}$ and $\widehat{HO_dO_a}$ are shown in Table 1. The first remark is that the average values obtained in all test simulations for the benchmark unconfined methane hydrate and bulk liquid water systems are in good agreement with theoretical values, supporting our AA-MD approach.

The hydrogen bond angle $\widehat{O_dHO_a}$ was 157–160° in the benchmark bulk water system whereas an angle of 167° was obtained for the benchmark bulk hydrate system, in good agreement with the theoretical range of 130–180°[52]. The $\widehat{HO_dO_a}$ displayed angles between 12–14° for benchmark bulk water system and 8–9° for benchmark bulk hydrate system. Depending on the water molecule arrangement, differences of approximately 7–10° in the $\widehat{O_dHO_a}$, and 4–6° in the $\widehat{HO_dO_a}$ angles were found. As noticed in $F_3$ and $F_4$ order parameters, the results obtained for these hydrogen bond angles seem to be independent of the thermodynamic conditions at which the simulations were performed.

Then, with the objective of characterising the structure of our confined system, the pore between the silica walls was split into 35 slabs parallel to the silica surface. The values of $F_3$ and $F_4$ order parameters, and hydrogen bond angles $\widehat{O_dHO_a}$ were averaged for each slab, and the profiles obtained along the simulation box are shown in Fig. 5 for the NVT run at 273 K Again, the corresponding plots for the other simulation runs performed have been included as in supporting information in the Appendix, evidencing the same behaviour. The $\widehat{HO_dO_a}$ was also computed, but for the sake of clarity it was not plotted, since it exhibits the same $\widehat{O_dHO_a}$ angle trend. The order parameters and hydrogen bond angle displayed different trends depending on the position inside the nanopore. As in the case of the density profiles, a layer of approximately 1.0 – 1.5 nm width next to both silica walls was clearly identified, where the parameters exhibited different values compared with those in the nanopore centre. Thus, the previous notation for these regions, H layer (hydrate), and F layer (adsorbed fluid) is used to discuss their distinctive features. The width of this fluid water layer is similar to that obtained previously by Conde et al. [56], who performed MD simulations for studying the thickness of a water liquid layer on the free surface of ice. They employed typical water models, including the TIP4P/Ice also employed in this work, and obtained a layer 1 nm width. Later, Shepherd et al. [57] studied the quasi-liquid water layer formed in the surface of ice under hydrate for-mation conditions, in the presence of methane and using a coarse–grained approach, obtaining a similar liquid layer thickness.

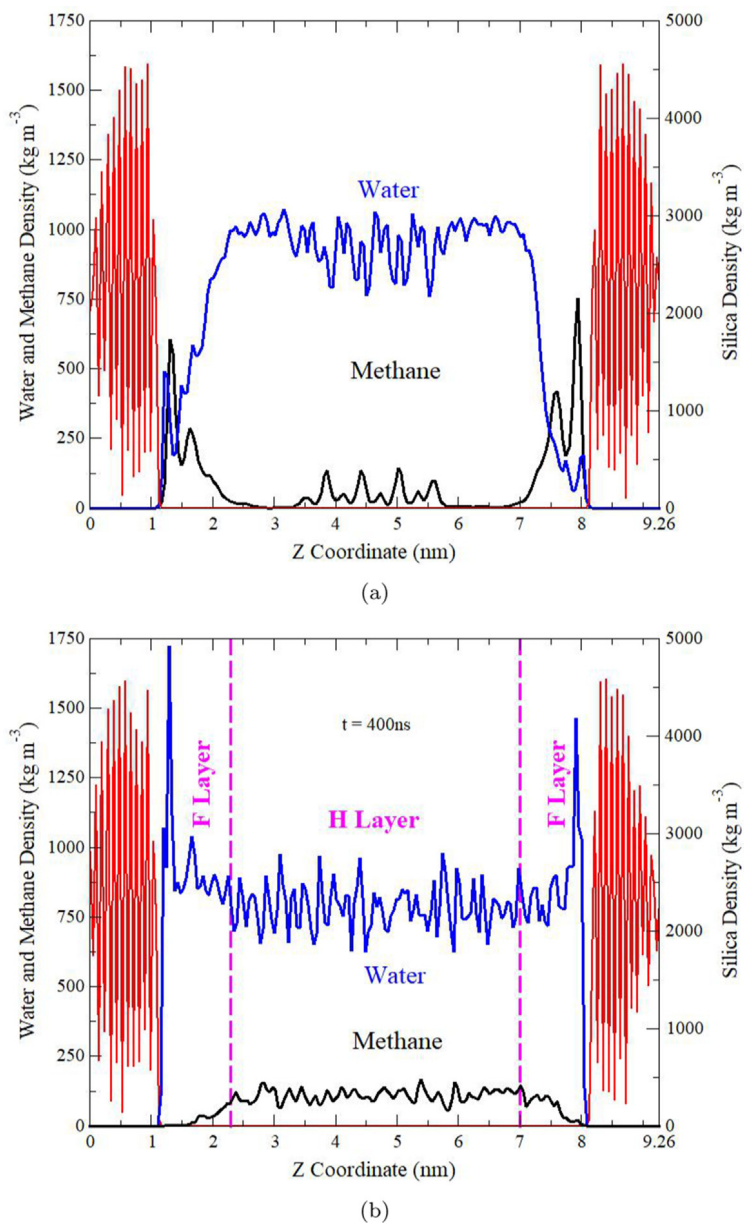

**Fig. 4.** Density profile along Z axis of silica wall (red), water (blue) and methane (black), determined at the beginning (4a) and the end (4b) of the simulation in the NVT ensemble at T = 273 K. The vertical dashed lines in Fig. 4b are shown as a guide to the eye to separate the central region occupied by the crystallized methane hydrate (H layer), and the fluid layers between the hydrate and the silica w.alls (F layers).

**Table 1**
Theoretical and computed $F_3$ and $F_4$ order parameters, and hydrogen bond angles, for hydrate and water phase.

| | | Order Parameters | | Hydrogen Bond Angle (degrees) | |
|---|---|---|---|---|---|
| | | $F_3$ | $F_4$ | $\widehat{O_d H O_a}$ | $\widehat{H O_d O_a}$ |
| Confined system H Layer (Hydrate) | NpT p = 100 bar T = 260 K | 0.022 | 0.611 | 166.46 | 8.98 |
| | NVT T = 260 K | 0.023 | 0.569 | 166.23 | 9.14 |
| | NVT T = 273 K | 0.017 | 0.689 | 166.42 | 9.01 |
| Benchmark Bulk Hydrate | NVT T = 260 K | 0.013 | 0.732 | 167.22 | 8.47 |
| | NVT T = 273 K | 0.013 | 0.722 | 166.90 | 8.68 |
| | NpT p = 100 bar T = 260 K | 0.013 | 0.731 | 167.20 | 8.48 |
| Benchmark Water | NpT p = 1 bar T = 300 K | 0.085 | −0.044 | 160.33 | 12.82 |
| | NpT p = 1 bar T = 350 K | 0.095 | −0.023 | 157.41 | 14.67 |
| Theoretical Values | Bulk Hydrate | 0.0 [48] 0.01 [51] | 0.7 [49,51] | 130–180 [52] | <30 [53] |
| | Ice | 0.01 [51] | −0.4 [51] | 130–180 [52] | <30 [53] |
| | Water | 0.1 [48,51] | 0.0 [49]-0.04 [51] | – | – |

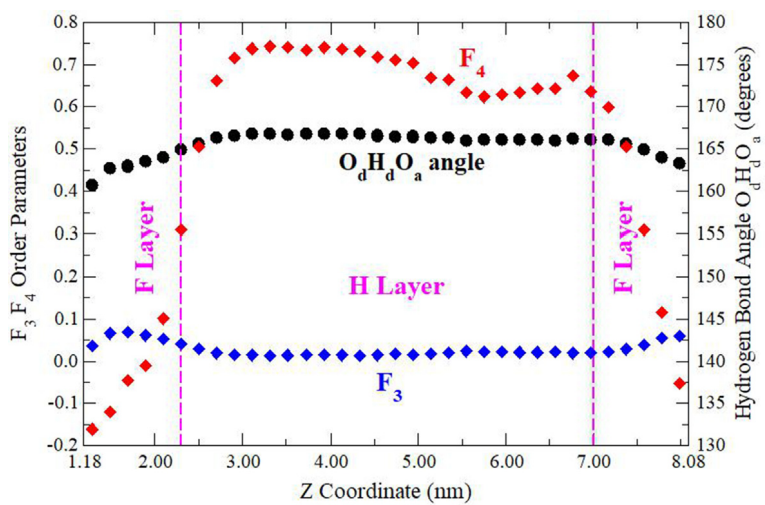

**Fig. 5.** Profiles of order parameters $F_3$ (◆) and $F_4$ (◆), and hydrogen bond angle (●) along Z axis were taken from the NVT run .at T = 273 K.

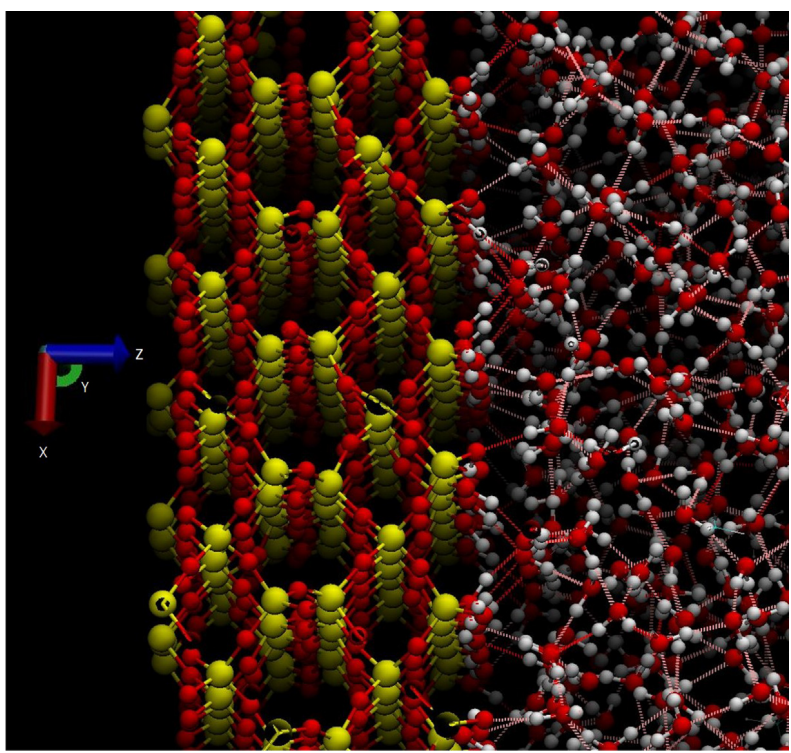

**Fig. 6.** Magnification of Fig. 2 highlighting the immediacy of the silica wall, showing the formation of hydrogen bonds between water molecules and hydroxyl groups of the silica wall. The oxygen atoms of the water molecules act as proton acceptor, and the hydrogen atom of the silica surface as donor. On the contrary, the oxygen atoms of the hydroxyl groups act as acceptor of the hydrogen .water molecules.

In addition to the profiles shown in Fig. 5, the average values of the determined parameters are also displayed in Table 1. The value obtained for the $F_3$ order parameter in the H layer is in very good agreement with the theoretical one as well as with those obtained in our benchmark simulations performed for a type *I* hydrate structure. The above results were also in concordance with the simulations performed by Fang et al. [19] who stated a $F_3$ value below 0.05 as hydrate identification criterion. This supports our previous statement regarding the structure grown inside the nanopore is methane hydrate. In this H layer, the $F_4$ order parameter also yields an average value in very good agreement with hydrate references, and it was also in agreement with the work of Walsh et al. [50] who studied the nucleation and growth of methane hydrate, monitoring the $F_4$ order parameter until a value of 0.6 was reached at the end of their simulations. Thus, the $F_3$ and $F_4$ parameter profiles allow to identify unambiguously the solid grown within the pore as a hydrate.

Concerning the adsorbed F layer, $F_3$ order parameter increased until $\approx 0.06$. This is not exactly the theoretically expected value for water (0.1) but this increase means that water molecules do not show tetrahedral arrangement in this layer. In the slabs in contact with the silica walls, $F_3$ decreased again to 0.04, meaning that water molecules in contact with silica wall recover some ordering. This can be explained by the fact that these water mole-cules in contact with the silica wall established hydrogen bonds with the silica wall terminal hydroxyl groups. This hypothesis can be verified in Fig. 6, where a magnification of Fig. 2 is presented. Fig. 6 shows the interfacial region between a silica wall and the F layer, where this crossed association phenomenon is clearly depicted. Additionally, in F layer, the $F_4$ decreases drastically from 0.6 to −0.15 in the silica wall boundary. This trend clearly denotes that water molecules no longer show hydrate structure, but a disordered fluid-like structure instead. The $F_4$ values in the water slab in contact with the silica wall confirmed the formation of hydrogen bonds with silica hydroxylated groups, and thus these water molecules displayed a higher ordering (although still fluid).

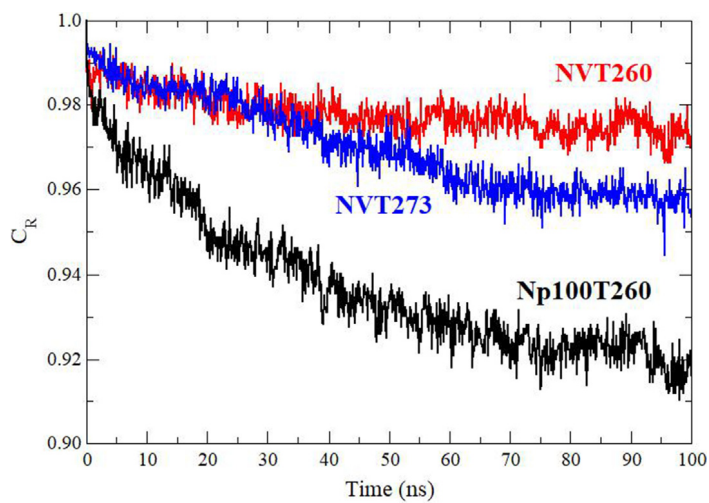

**Fig. 7.** Correlation function $C_R$ for water molecules in the F layer, represented along the production stage for the three simulations of the confined systems (Ta.ble 1).

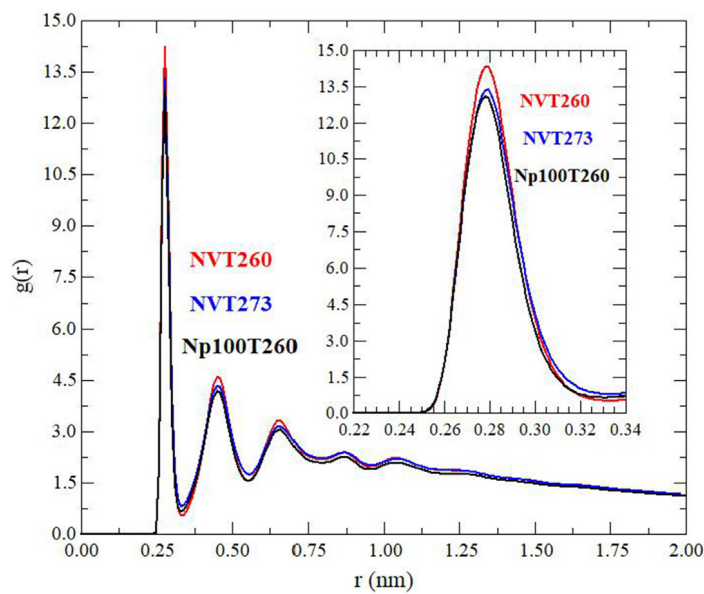

**Fig. 8.** Radial distribution function for water molecules in the F layer, for the three simulations performed.

Finally, the hydrogen bond angles $\widehat{O_d H O_a}$ and $\widehat{H O_d O_a}$ showed a compatible behaviour with the order parameters in this F layer. Their values drop from 166 to 161° for $\widehat{O_d H O_a}$ angle and from 9 to 11° for $\widehat{H O_d O_a}$ angle.

Water molecules in F layer did not exhibit hydrate structure, as shown by the $F_3$ and $F_4$ parameter profiles. With the aim to shed light into the nature of this F layer, we have computed the residence correlation function $C_R$ for those water molecules. As shown in Fig. 7, more than 90% of water molecules remained in this layer after 100 ns. This means that the exchange of water molecules between this layer and the hydrate is very limited and were preferentially arranged at the silica surface. Additionally, the radial distribution functions (RDF) were obtained for the water molecules in this F layer, and are displayed in Fig. 8. The pattern found in the RDF for the F layer corresponds to a typically fluid phase, discarding the possibility of a solid arrangement. The Mean Square Displacement (MSD) was also obtained and displayed in Fig. 9. The diffusion coefficient can be computed from the slope of this figure. The values obtained were $1.97 \cdot 10^{-12}$, $10.36 \cdot 10^{-12}$ and $19.61 \cdot 10^{-12} m^2/s$ in the NVT (T = 260 K), NpT (p = 100 bar and T = 260 K) and in the NVT (T = 273 K) ensembles, respectively. These results evidenced that water molecules do diffuse across the F layer, confirming the hypothesis of a fluid phase, and discarding the possibility of any type of glassy state. It must be pointed out that our diffusion coefficients are in the same order of magnitude than those obtained by *Cao et al.* [58] for the diffusion of $H_2$ through clathrate hydrates, and also for those of the diffusion of $H_2$ through hydroquinone clathrates [25].

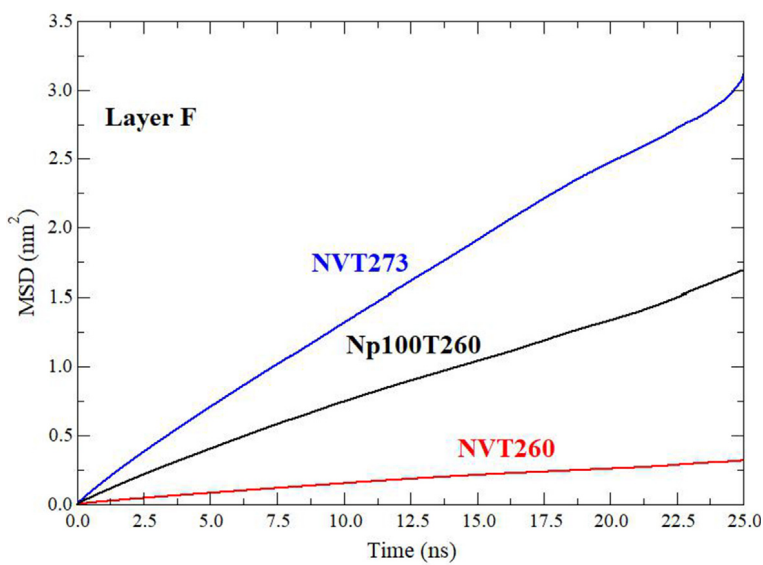

**Fig. 9.** Mean square displacement (MSD) for water molecules in the F layer for the three simulations of the confined system.

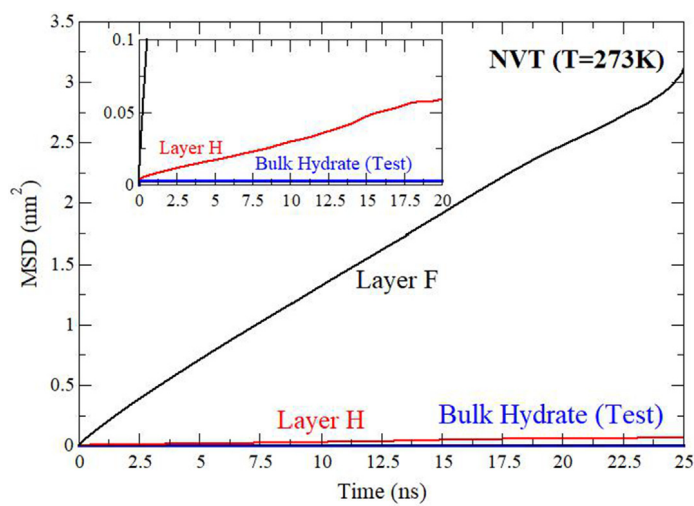

**Fig. 10.** MSD for F (black line) and H (red line) layers in the NVT run at 273 K. The MSD for a perfect methane hydrate crystal is represented in blue for the same conditions.

Finally, a comparison the MSD of water molecules in both F and H layers is represented in Fig. 10 for the simulation performed in the NVT ensemble at 273 K. The trend of an unconfined hydrate is also represented as a reference. As it can be seen, the H layer, containing the hydrate structure in our confined system, displayed a very similar water mobility than the unconfined system. The fact that the diffusion coefficient of the H layer was somewhat higher, provided an additional evidence of the presence of defects in the confined hydrate structure, likely due to the constraints inherent to the spatial confinement. The existence of these imperfections resulted in a certain (although vanishingly small) possibility of water molecule displacements, when compared with a perfectly regular hydrate structure. The water diffusion coefficient in the F layer was clearly higher than those observed in the benchmark bulk hydrate and benchmark water system shown in Table 1. The same behavior can be seen for the other simulations, whose figures have been included in the Appendix.

## 4. Conclusions

The structural and thermodynamic characterisation of bulk or unconfined gas hydrates using different molecular simulation techniques has been subject of many studies in scientific literature, providing insightful understanding regarding the behaviour of these inclusion solids. In contrast, the hydrate behaviour in confined geometries has been by far less studied despite the fact that natural hydrate deposits are often found in sedimentary substrates of variable porosity. Thereby, determining the influence of these solid substrates on the hydrate structures, and the impact in their thermophysical, equilibrium and transport properties is a challenging objective.

With this objective, Molecular Dynamics has been used to study the crystallisation of methane hydrate within a slit silica pore. The silica α-quartz crystalline structure has been built with a fully atomistic force field, describing all intermolecular interactions to simulate a flexible structure, in contrast with previous computational approaches found in literature. Thus, a slit pore of roughly 7 nm width has been constructed. Inside the pore, the initial simulated system contained a seed of methane type *I* hydrate, liquid water, and methane. The system was then setup in thermodynamic conditions of hydrate stability, performing both NVT and NpT simulation. After the energy minimisation, the system evolved in time with the initial hydrate seed growing to occupy all the accessible volume within the pore. In all simulation runs performed the final structure inside the pore was virtually the same.

The geometrical constraint imposed by the presence of the solid walls did not prevent hydrate crystallisation. The guest methane molecules gradually diffused towards the hydrate interface, and liquid water molecules joined the hydrate crystalline structure growing it, yielding a final slab of type *I* hydrate, but presenting noticeable imperfections and defects. The appearance of these defects was caused because the available space for the hydrate growth is not a multiple of the basic hydrate crystalline cell dimensions, and thus the growing hydrate accommodated to occupy the accessible volume. The water molecules which were not incorporated to the hydrate structure formed, at the end of the simulation, a confined fluid layer between the hydrate and the silica walls.

The nature of the adsorbed water layers (denoted previously as F layers) was then analyzed in detail. The anomalous high density observed could point towards a glassy solid state. Definitely it is not another crystalline phase, as the absence of order is clear. Trying to find support for its identification as a fluid phase, the correlation function indicating residence time, the radial distribution function, and the MSD of the water molecules within this F layer were computed. All of them exhibited clear trends of fluid state behaviour. The same was observed analysing the profile of hydrogen bond angles, providing an additional interesting hint. In this regard, water molecules in contact with the silica wall established hydrogen bonds with the terminal hydroxyl groups of silica molecules, resulting strongly attached to the silica hydrophilic surface.

The scenario resulting from our molecular simulation calculations offers a enlightening description of hydrate crystallisation process under confinement. Our study analysed the impact of distinctive features as the consequence of numerous defects and a distorted albeit recognisable structure, as well as the presence of an interstitial adsorbed fluid water layer between the hydrate grown and the confining solid substrate. At the same time, many engaging questions remain, as the difference of guest occupation or mobility, or the probable shift in hydrate phase boundaries due to confinement. In the progress towards answers to these questions, the molecular simulation setup used in this work has shown to be an extremely useful theoretical approach, providing a versatile computer simulation framework to evaluate the impact of confinement conditions in hydrate growth.

**Declaration of Competing Interest**

The authors declare that they have no known competing financial interests or personal relationships that could have appeared to influence the work reported in this paper.

**Acknowledgements**

A.M.F.F., M.P.R. and M.M.P. acknowledge funding by Ministerio de Ciencia e Innovación [Grant No. PID2021-125081NB-I00] cofinanced by the European Regional Development Fund, ERDF, Sudoe EU Program KET4F-Gas [Grant No. SOE2/P1/P0823], and Consellería de Educación Universidade e Formación Profesional, Xunta de Galicia [FSE-GALICIA 2014–2020]. G.P.S. declares that this work was developed within the scope of the project CICECO-Aveiro Instituto de Materiais, UIDB/50011/2020, UIDP/50011/2020 & LA/P/0006/2020, financed by national funds through the FCT/MEC (PIDDAC), and also acknowledges national funds (OE), through FCT – Fundação para a Ciência e a Tecnologia, I.P., in the scope of the framework contract foreseen in the numbers 4, 5 and 6 of the article 23, of the Decree-Law 57/2016, of August 29th, changed by Law 57/2017, of July 19th. M.M.C. acknowledges financial support from the MICINN (Grant No. PID2019-105898GA-C22) and CAM and UPM through the Cavities (Project No. APOYO-JOVENES-01HQ1S-129-B5E4MM) from "Acción financiada por la Comunidad de Madrid en el marco del Convenio Plurianual con la Universidad Politécnica de Madrid en la linea de actuación estímulo a la investigación de jóvenes doctores." The authors also gratefully acknowledge computing resources provided by Universidad Politécnica de Madrid (www.upm.es, Magerit Supercomputer) and Centro de Supercomputación de Galicia (CESGA, www.cesga.es), and Universidade de Vigo/CISUG for open access charge funding.

**Appendix A. Additional Figures**

Figs. 11–13

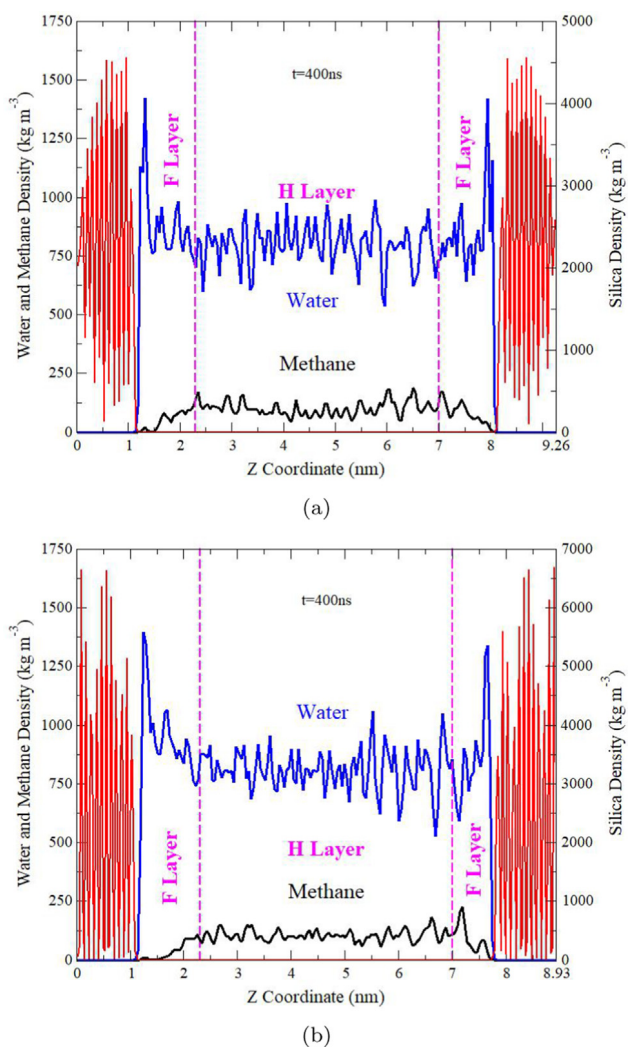

**Fig. 11.** Density profile along Z axis of silica wall (red), water (blue) and methane (black), determined at the end of the simulation. Fig. 11 shows the results in the NVT ensemble at T = 260 K and Fig. 11b in the NpT ensemble at p = 100 bar and T = 260 K. The vertical dashed lines are shown as a guide to the eye to separate the central region occupied by the crystallized methane hydrate (H layer), and the liquid layers between the hydrate and the silica w.alls (F layers).

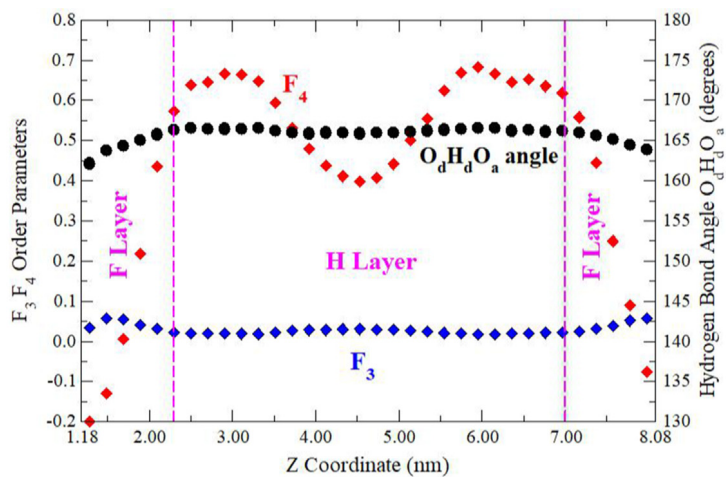

(a)

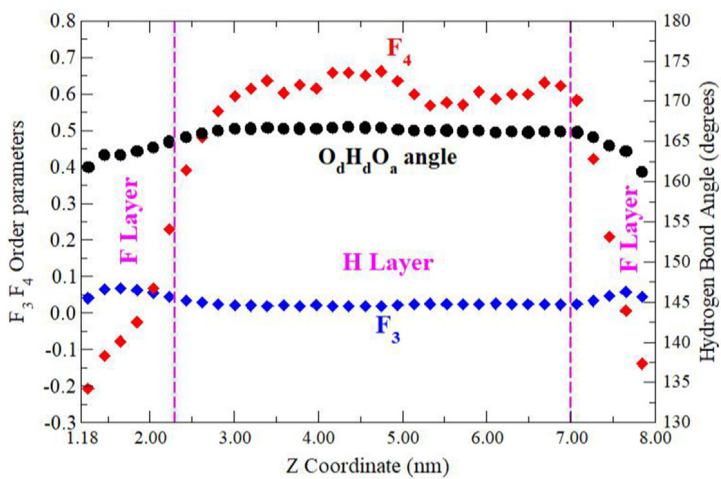

(b)

**Fig. 12.** Profiles of order parameters $F_3$ (◆) and $F_4$ (◆), and hydrogen bond angle (●) along Z axis. Fig. 12a shows the profile for the NVT ensemble at 260 K and Fig. 12b for the NpT ensemble at 100;bar and 260 K.

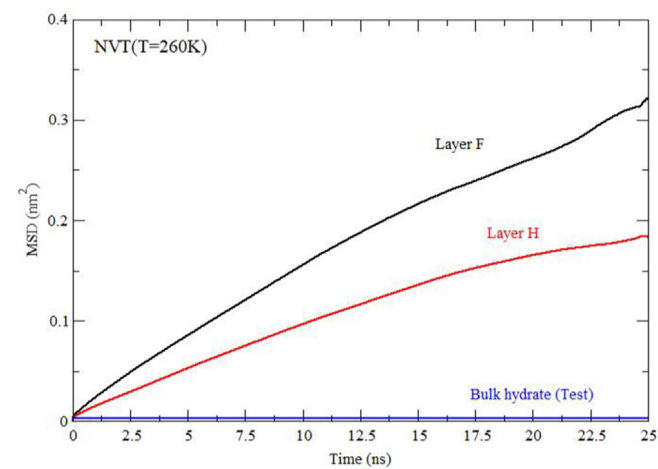

(a)

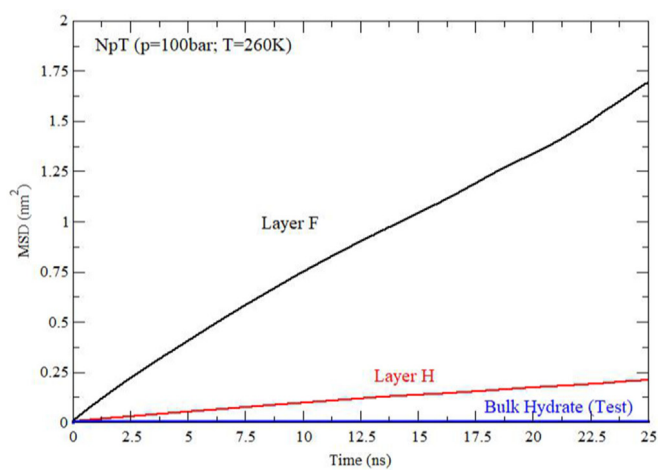

(b)

**Fig. 13.** Mean square displacement for F (black line) and H (red line) layers in the NVT (Fig. 13a) and in the NpT (Fig. 13b) ensembles. Both were performed at 260 K and in the case of the NpT ensemble, the barostat was set up at 100 bar. MSD is also plotted in blue for the same conditio.ns in each case.